\documentclass[aps,prl,twocolumn,showpacs,groupedaddress,floatfix,
               amsmath,amssymb]{revtex4}

\usepackage{graphicx}%
\usepackage{dcolumn}
\usepackage{bm}

\def\urltilda{\kern -.15em\lower .7ex\hbox{\~{}}\kern .04em}

\begin{document}

\title{Efficient Computation of Casimir Interactions between Arbitrary 3D Objects}

\author{M. T. Homer Reid$^{1,2}$%
        \footnote{Electronic address: \texttt{homereid@mit.edu}}%
        \footnote{URL: \texttt{http://www.mit.edu/\urltilda homereid}},
       Alejandro W. Rodriguez$^1$, 
       Jacob White$^{2,3}$,
       and Steven G. Johnson$^{2,4}$}
\affiliation{ $^{1}$Department Of Physics, 
              Massachusetts Institute of Technology, 
              Cambridge, MA 02139, USA \\
              $^{2}$Research Laboratory of Electronics,
              Massachusetts Institute of Technology, 
              Cambridge, MA 02139, USA \\
              $^{3}$Department of Electrical Engineering and Computer Science,
              Massachusetts Institute of Technology, 
              Cambridge, MA 02139, USA \\
              $^{4}$Department Of Mathematics,
              Massachusetts Institute of Technology, 
              Cambridge, MA 02139, USA}
\date{\today}

\begin{abstract}
We introduce an efficient technique for computing Casimir energies 
and forces between objects of arbitrarily complex 3D geometries.
In contrast to other recently developed methods, our technique
easily handles non-spheroidal, non-axisymmetric objects and 
objects with sharp corners. Using our new technique, we obtain the
first predictions of Casimir interactions in a number of experimentally 
relevant geometries, including crossed cylinders and tetrahedral 
nanoparticles.
\end{abstract}

\maketitle

Since the dawn of the modern era of
precision Casimir-force measurements some 10 years 
ago~\cite{precision}, Casimir forces
have been measured in an increasingly wide variety of 
experimental geometries, including 
plate-plate~\cite{plateplate}, sphere-plate~\cite{sphereplate}, 
sphere-comb~\cite{spherecomb}, and cylinder-cylinder~\cite{cylcyl} 
configurations. A recent experiment~\cite{raffaele} finds evidence
of Casimir interactions in a commercially fabricated MEMS device, 
and there is every reason to believe that accurate modeling of Casimir 
forces in complex geometries will be a critical ingredient in the 
design of future commercial MEMS technologies~\cite{ieee1,ieee2,ieee3}.

Theoretical developments have largely failed to keep pace with this 
rapidly advancing experimental reality. Until recently, theoretical
considerations were restricted to simple models in highly idealized 
geometries~\cite{casimir}, such as infinite parallel plates or 
infinite parallel cylinders. Techniques for predicting Casimir forces 
in more general geometries are clearly needed if theory is to confront 
the growing 
wealth of experimental Casimir data and the design challenges of future 
MEMS devices. 

Important recent progress on general methods has been made by a number of 
authors~\cite{alex1,EGJK,PasqualiMaggs1,Lambrecht08,Gies06}.
Rodriguez \textit{et. al.}~\cite{alex1} applied standard techniques of 
computational electromagnetics to develop a method capable of predicting 
Casimir forces between arbitrary \textit{two-dimensional} objects, 
i.e.~infinitely extended objects of arbitrary 2D cross section. 
Emig \textit{et. al.} (EGJK)~\cite{EGJK} demonstrated an efficient 
algorithm for predicting Casimir energies of interaction between compact 
3D objects of \textit{spheroidal} or \textit{nearly-spheroidal} shape. 
Both of these techniques have since been applied to predictions of Casimir 
interactions in new geometries~\cite{alex2,alex3,EGJK2}. To date, however, 
a practical computational scheme for predicting Casimir forces between 
arbitrary non-spheroidal three-dimensional objects has remained elusive.

\begin{figure}[b]
\includegraphics{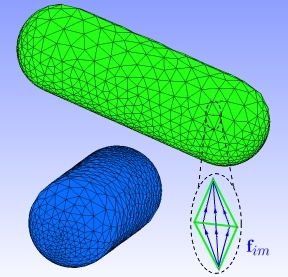}
\caption{
Surfaces of compact conducting objects are 
discretized~\cite{GMSH} into planar triangles, which
are used to construct localized vector-valued basis 
functions describing the surface current distribution.
As shown 
in the inset, each basis function describes currents flowing 
from a vertex of one triangle to the opposite vertex of an 
adjacent triangle, and vanishes on all other triangles~\cite{RWG}.
}
\label{RWGfig}
\end{figure}

In this work, inspired by the ideas of Refs.~\onlinecite{alex1} 
and~\onlinecite{EGJK}, we introduce a new technique for calculating 
Casimir forces between objects of arbitrary 3D shapes, including 
efficient handling of 
non-spheroidal, non-axisymmetric objects and objects with 
sharp corners. As an immediate application of our method, we present 
the first predictions of Casimir interactions in a number of 
experimentanlly relevant geometries, including crossed
cylindrical capsules~\cite{cylcyl} and tetrahedral
nanoparticles~\cite{Tetrahedra1,Tetrahedra2}.

\textit{The method.} The starting point of the EGJK method is a
path-integral expression~\cite{LiKardar} for the Casimir energy 
of perfectly-conducting compact objects: 
\begin{equation}
\mathcal{E}=-\frac{\hbar c}{2\pi} \int_0^\infty d\kappa 
   \log\frac{\mathcal{Z}(\kappa)}{\mathcal{Z}_\infty(\kappa)},
\label{Eexp}
\end{equation}
\vspace{-0.2in}
\begin{equation}
\mathcal{Z(\kappa)}
=\int \mathcal{D} \mathbf J(\mathbf x)\,
 e^{  -\frac{1}{2}\int d\mathbf x \, \int \, d\mathbf x^\prime \, 
       \mathbf{J}(\mathbf x) \cdot 
       \bm{\mathcal G}_\kappa (\mathbf x, \mathbf x^\prime)
       \cdot \mathbf{J}(\mathbf x^\prime) 
    }
\end{equation}
where the functional integration extends over all possible current
distributions $\mathbf J(\mathbf x)$ on the surfaces of the objects 
and where 
$ \bm{\mathcal{G}}_\kappa=
  \left[ \mathbf{1}+\frac{1}{\kappa^2}\nabla\otimes\nabla^\prime\right]
         \frac{e^{-\kappa|\mathbf{x}-\mathbf{x}^\prime|}}
              {4\pi|\mathbf{x}-\mathbf{x^\prime}|}$
is the dyadic Green's function at (Wick-rotated) frequency 
$-i\omega=c\kappa$.
$\mathcal{Z}_\infty$ in (\ref{Eexp}) is $\mathcal{Z}$ computed
with all objects removed to infinite separation. 

Following EGJK, to obtain a tractable expression for $\mathcal{Z}$ for a 
given collection of $N_o$ objects we now proceed to expand the current 
distribution in a discrete set of basis functions,
$\mathbf J(\mathbf x)=\sum J_{im} \mathbf f_{im}(\mathbf x),$ 
where $i=1\cdots N_o$ ranges over the objects in the geometry
and $m=1\cdots N_i$ runs over a set of $N_i$ expansion functions 
defined for the $i$th object. With this expansion we have
\begin{eqnarray}
 \mathcal{Z}(\kappa)
   &=&\mathcal{J} \int \prod dJ_{im} 
                 e^{-\frac{1}{2}\sum J_{im}\mathbf{M}^{(\kappa)}_{im,jn} J_{jn} }
       \label{gi} \\
   &=&(2\pi)^{n/2}\mathcal{J} \cdot \left[\det \mathbf M(\kappa)\right]^{-1/2}
\end{eqnarray}
and (\ref{Eexp}) becomes
\begin{equation}
  \mathcal{E}=+\frac{\hbar c}{2\pi}\int_0^\infty d\kappa 
   \,\log \frac{\det \mathbf{M}(\kappa)}{\det \mathbf{M}_\infty(\kappa)}
  \label{Eexp2}
\end{equation}
where we have evaluated the multiple Gaussian integral 
in (\ref{gi}) using standard path-integration techniques~\cite{Weinberg}.
Here the elements of the matrix $\mathbf M(\kappa)$ are the interactions 
between the basis functions, 
$\mathbf M_{\alpha\beta}(\kappa)=
 \iint \mathbf{f}_{\alpha}\cdot \bm{\mathcal{G}_\kappa} 
 \cdot \mathbf{f}_{\beta}\, d\mathbf x \,d\mathbf x^\prime
$,
and $\mathcal{J}$ in (\ref{gi}) is the (constant) Jacobian of the 
transformation $\mathcal{D}\mathbf J\to \prod dJ_{im}$, which cancels in 
the ratio (\ref{Eexp2}). 

EGJK took the expansion functions $\mathbf{f}_{im}$ 
to be proportional to the spherical multipole moments 
$\{Q^i_{\scriptsize{E},lm},Q^i_{\scriptsize{M},lm}\}$ of the 
$i$th object. Although this choice leads to rapidly convergent
and even analytically tractable series for \textit{spherical} 
objects, it is not of practical use for general geometries, 
as the spherical multipole expansion must be carried to high
orders to represent source distributions on highly 
non-spherical objects. Ref.~\onlinecite{ShihHsien}, for example, 
demonstrated poor convergence rates for spherical-harmonic 
representations of elongated objects and objects with corners 
or cusps.

A more general strategy is to discretize the surfaces of the 
objects into planar triangles, as depicted in Figure \ref{RWGfig}, 
and to introduce \textit{localized} basis functions in the spirit
of finite-element and boundary-element methods. Following
standard practice in computational electromagnetics~\cite{RWG}, 
we choose basis functions defined on pairs of adjacent 
triangles. As indicated in the inset of Figure 1, to the $m$th 
internal edge in the surface discretization of object $i$ we 
associate a localized basis function $\mathbf f_{im},$ which 
describes currents flowing on the two triangles that share
that edge and vanishes on all other triangles.
The matrix elements $\mathbf M_{\alpha\beta}$ reduce 
to finite surface integrals over plane triangles, the evaluation
of which is greatly facilitated by the elegant methods of 
Ref.~\onlinecite{DJTaylor}, and the overall accuracy of 
the computation may be systematically improved, at greater 
computational expense, by reducing the size of the triangles in 
the surface mesh. A further 
virtue of this choice of basis functions is that it eliminates 
the unwieldy distinction between interior and exterior fields 
in~\cite{EGJK}.

\textit{Casimir Energy.} To compute the energy of interaction of our
objects we write
\begin{equation}
 \log\frac{\det\mathbf M}{\det\mathbf M_\infty}=
 \sum_{n=1}^N 
  \left\{\vphantom{\prod}\log \lambda_n - \log\lambda_n^\infty\right\}
 \label{eigenergy}
\end{equation}
where $\{\lambda_n\},\{\lambda_n^\infty\}$ are the eigenvalues
of $\mathbf{M},\mathbf{M}_\infty.$ 

\textit{Casimir Force.} The $z$-directed force on the $i$th object 
in our geometry is obtained by differentiating (\ref{Eexp2}) with
respect to displacement of that object:
\begin{equation}
\mathcal{F}_{z,i}=-\frac{\hbar c}{2\pi} \int_0^\infty d\kappa \,
 \frac{\partial }{\partial z_i} \ln 
 \frac{\det{\mathbf{M}(\kappa)}}{\det{\mathbf{M}_\infty(\kappa)}}.
 \label{forceintegral}
\end{equation}
The integrand may be conveniently evaluated using the 
identity~\cite{EmigEurophys}
$$ \frac{\partial}{\partial z_i} \ln \det{\mathbf{M}}
  =\text{Tr} 
   \left\{ \mathbf M^{-1}
           \cdot 
           \frac{\partial \mathbf M}{\partial z_i} 
  \right\}
  =\sum_{n=1}^N \alpha_n 
$$
where $\{\alpha_n\}$ are the eigenvalues of the \textit{generalized}
eigenvalue problem
\begin{equation}
 \frac{\partial \mathbf M}{\partial z_i} \cdot \mathbf{v} = 
   \alpha \mathbf{M}\cdot \mathbf{v}.
  \label{eigforce}
\end{equation}
The $\kappa$ integrals in (\ref{Eexp2}) and (\ref{forceintegral}) may 
be evaluated using standard numerical quadrature techniques.

The dimension of the matrix $\mathbf{M}$ is the number of basis functions 
in our expansion of the surface current distribution, which in turn is
proportional to the number of triangles in our surface discretization
(see Figure \ref{RWGfig}).
For matrices of moderate dimension, $N \lesssim 5000$, corresponding to 
moderately fine mesh discretizations (sufficient for accurate treatment
of all geometries considered here), the eigenvalues in 
(\ref{eigenergy}) and (\ref{eigforce}) may be computed using
direct methods in $\mathcal{O}(N^3)$ time. For finer meshes it may be 
possible to use iterative eigenvalue solvers and matrix sparsification 
techniques~\cite{ZZhu,Rokhlin85,Hackbusch89} to reduce the complexity 
to $\mathcal{O}(N\ln N);$ this will be discussed in a subsequent publication.

\begin{figure}
\includegraphics{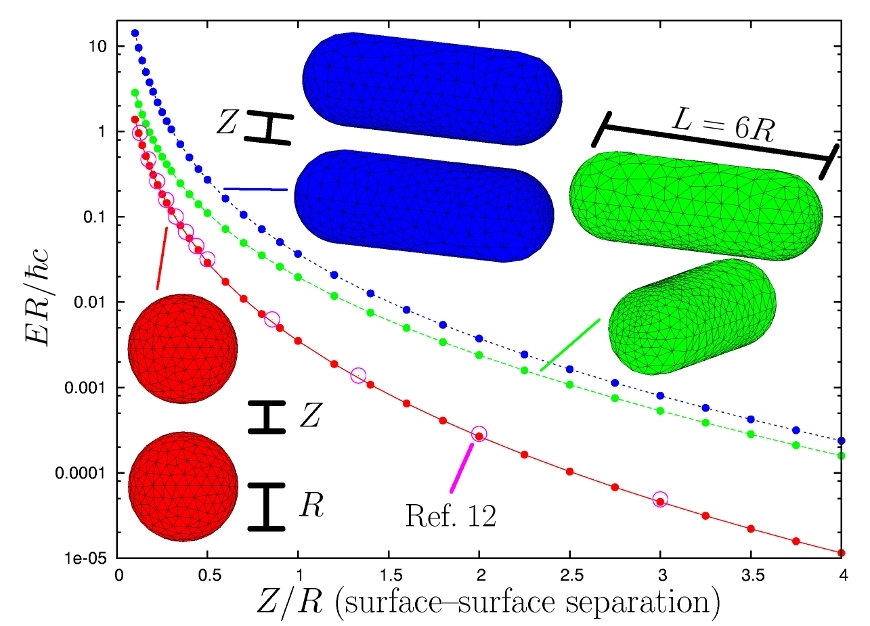}
\caption{Casimir energy vs. separation distance for three pairs 
of perfectly conducting objects: identical spheres (red curve), 
identical capsules with parallel axes (blue curve), and identical 
capsules with perpendicular axes (green curve). The hollow
circles represent the sphere-sphere data of Ref.~\onlinecite{EGJK}.
}
\label{Figure2}
\end{figure}

\begin{figure}
\resizebox{8.6cm}{!}{\includegraphics{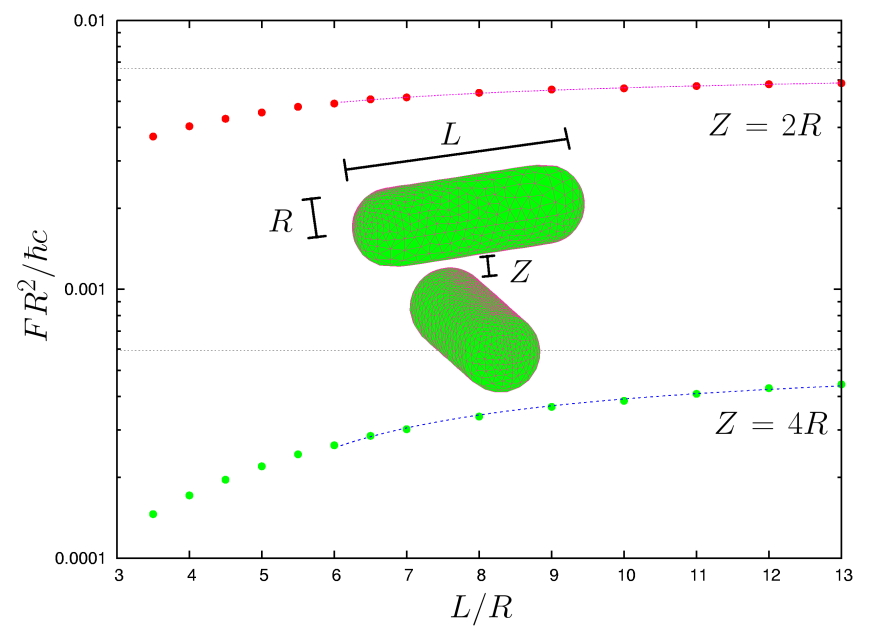}}
\caption{Magnitude of attractive Casimir force between crossed
capsules of radius $R$ as a function of capsule length, for 
surface--surface separations $Z=2R$ and $Z=4R$. The dashed
lines denote the asymptotic $(L\to\infty)$ limits of the 
force, as roughly extrapolated from our finite-$L$ data. The 
solid lines are fits of the large-$L$ data to the form $a+b/L$, 
although our data are insufficient to establish
the precise asymptotic $L$-dependence of the force.}
\label{Figure3}
\end{figure}

\begin{figure*}
\resizebox{15cm}{!}{\includegraphics{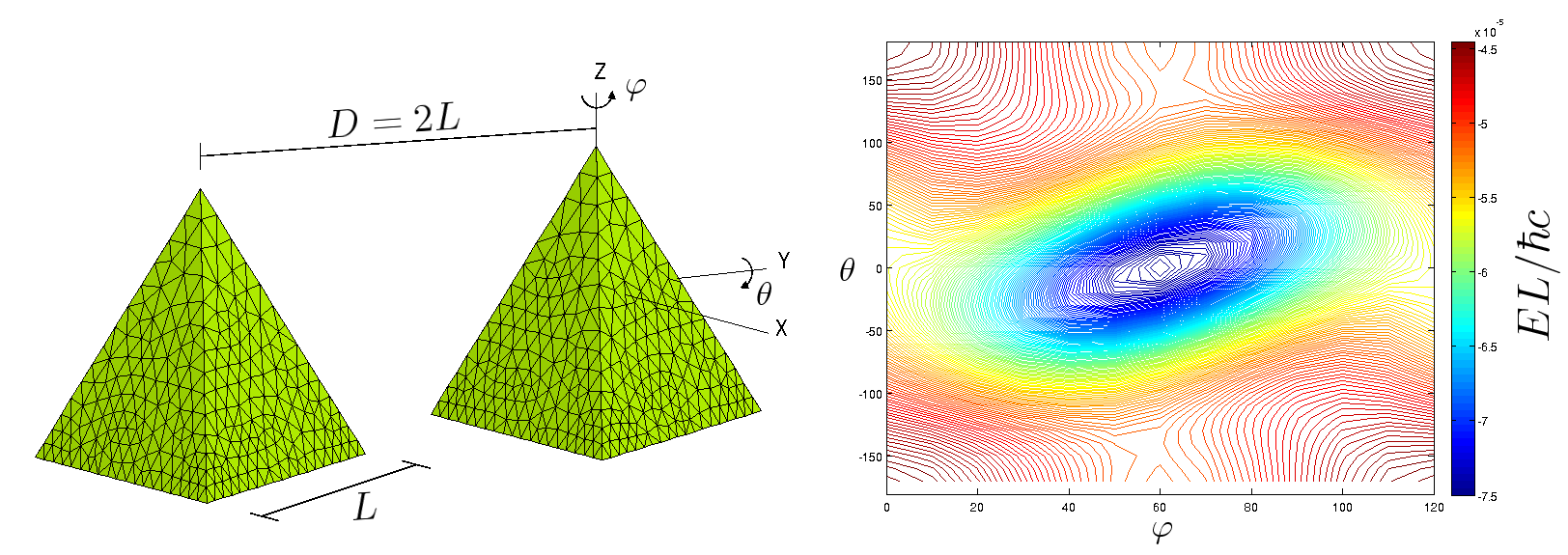}}
\caption{Contour plot of Casimir energy vs. orientation angles 
for tetrahedral nanoparticles separated by a distance $D=2L.$}
\label{Figure4}
\end{figure*}

\textit{Results.} We now apply our technique to the prediction 
of Casimir interactions in a number of 3D geometries. 

\textit{Casimir energy of identical spheres, parallel capsules, and 
crossed capsules.} Figure 2 shows plots of Casimir energy versus 
separation distance for three pairs of perfectly conducting objects: 
identical spheres of radius $R$, identical capsules with parallel axes, 
and identical capsules with perpendicular axes (``crossed capsules.'') 
(A ``capsule'' is a cylinder of radius $R$ with hemispherical endcaps, 
of total length $L=6R$ in this case.) The energy curve for the crossed 
capsules interpolates between the curve for the spheres at short distances and 
the curve for the parallel capsules at large distances. We may understand 
this result as follows: At short 
distances, the interaction is dominated by contributions from 
the portions of the surfaces that lie in closest proximity to each other.
For the crossed capsules, this region of nearest proximity is restricted
to a length $\sim R$ around the centers of the capsules and hence
exhibits the same scaling as the region of nearest proximity 
for the two spheres in the short-distance limit. In contrast, for
parallel capsules, the strongly-interacting region extends down the 
entire length of the capsule and hence scales with an
additional factor of $L$ at short distances. At large distances, 
the interaction becomes pointlike and the energy of interaction
between identical capsules becomes independent of the orientation.

\textit{Casimir Force vs. Length For Crossed Capsules.} 
The Casimir force between crossed cylinders, in the limit 
in which the length $L$ of the cylinders is much longer than their
surface--surface separation $Z$, is experimentally 
accessible~\cite{cylcyl} and interesting in its own right as a 
finite interaction between effectively infinite objects. 
Figure 3 plots the magnitude of the attractive Casimir force between 
crossed metallic capsules of radius $R$, at fixed surface--surface 
separations of $Z=2R$ and $Z=4R$, as a function of the capsule length 
$L$. In the limit $L\to\infty$ our capsules become infinite cylinders
and the force approaches a $Z$-dependent constant (denoted by the dashed 
lines in the figure.)

An interesting question is how rapidly the force approaches the 
infinite-cylinder limit as $L\to\infty$ at fixed $Z.$ A numerical
determination of the precise asymptotic $L$-dependence of the force 
requires an iterative fast-solver version of our method 
that is capable of handling matrices $\mathbf M$ of 
larger dimension; this extension will be discussed in a future publication.

\textit{Tetrahedral nanoparticles.} Several groups have succeeded in 
producing tetrahedral nanoparticles of dimensions $\sim$ 10--50 
nm~\cite{Tetrahedra1,Tetrahedra2}, and Casimir forces may 
dominate the interactions between electrostatically neutral 
particles of this type. Our method allows the first predictions
of Casimir interactions between perfectly conducting tetrahedral 
particles. (In a real system with particles of these sizes, the finite 
conductivity of the metals cannot be neglected. We are currently 
implementing an extension of our method to treat objects of arbitrary 
permittivities. Even for the idealization of perfectly 
conducting nanoparticles, however, it is interesting to analyze the 
orientation dependence of the Casimir energy, as well as to demonstrate 
the practical ability of our method to treat geometries
with sharp corners and non-axisymmetric shapes.) Figure \ref{Figure4} 
depicts a contour plot of Casimir energy vs. orientation angles 
for two perfectly conducting tetrahedra originally separated by a distance 
$D=2L$ in the $y$ direction, where $L$ is the tetrahedron edge length. 
To one of the tetrahedra we apply a rotation through an angle $\varphi$ about 
the $z$ axis followed by a rotation through an angle $\theta$ about the $y$ 
axis; the fixed origin of these rotations, depicted as the origin of the 
coordinate axes in Figure \ref{Figure4}, is the point lying a distance 
$H/2$ below the apex of the tetrahedron, where $H=\sqrt{3}L/2$ is the 
height of the tetrahedron. The contour plot reveals a clear minimum at 
($\theta,\phi$)=(0,60$^\circ$), corresponding to the closest approach
of a vertex of the base of the rotated tetrahedron to the unrotated
tetrahedron, as might be expected for an attractive interaction.

\textit{Conclusion.} In conclusion, we have developed a general method 
for computing Casimir energies and forces between objects 
of arbitrary three-dimensional shapes, enabling efficient treatment 
of non-spheroidal, non-axisymmetric objects and objects with sharp 
corners. Using our method, we have predicted Casimir interactions 
in a variety of experimentally relevant geometries that would be
challenging to handle by previous methods.

\textit{Acknowledgements.}
We are grateful to T. Emig for providing the raw data from 
Ref.~\onlinecite{EGJK}.
The authors acknowledge the support of the Interconnect Focus Center, one
of five research centers funded under the Focus Center Research Program, a
DARPA and Semiconductor Research Corporation program. In addition, the
authors are grateful for support from the Singpore-MIT Alliance
Computational Engineering flagship research program. This work was supported
in part by a Department of Energy (DOE) Computational Science Fellowship
under grant DE-FG02-97ER25308 (ARW), by the Army Research Office
through the ISN under Contract No. W911NF-07-D-0004, and by the MIT 
Ferry Fund.


\end{document}